\documentclass[amsmath,amssymb,11pt,superscriptaddress,reprint, preprintnumbers, notitlepage,aps,prl,twocolumn]{revtex4-1}

\pdfoutput=1 
\usepackage[utf8]{inputenc}
\usepackage[english]{babel}
\usepackage{amsmath}
\usepackage{graphicx}
\usepackage{dcolumn}
\usepackage{pbox}
\usepackage{amssymb}
\usepackage{epsfig}
\usepackage[dvipsnames]{xcolor}
\usepackage{slashed}
\usepackage{amssymb}
\usepackage{ mathrsfs }
\usepackage{color}
\usepackage[font=small]{caption}
\usepackage[font=small]{subcaption}
\usepackage{url}
\usepackage[pdftex,bookmarks,linktocpage,pdfpagelabels,plainpages=false,hyperfigures,linkcolor=blue,citecolor=blue]{hyperref} 
\hypersetup{colorlinks=true}

\makeatletter

\makeatletter
\renewcommand\@makecaption[2]{%
  \par
  \vskip\abovecaptionskip
  \begingroup
  
   \small\rmfamily
    \begingroup
     \samepage
     \flushing
     \let\footnote\@footnotemark@gobble
     \@make@capt@title{#1}{#2}\par
    \endgroup
  \endgroup
  \vskip\belowcaptionskip
}
\makeatother

\DeclareUnicodeCharacter{2212}{-}
\begin{document}

\author{Vedran~Brdar}
\email{vedran.brdar@northwestern.edu}
\affiliation{Fermi National Accelerator Laboratory, Batavia, IL, 60510, USA}
\affiliation{Northwestern University, Department of Physics \& Astronomy, 2145 Sheridan Road, Evanston, IL 60208, USA}

\author{Sudip~Jana}
\email{sudip.jana@mpi-hd.mpg.de}
\affiliation{Max-Planck-Institut f\"ur Kernphysik  (MPIK), Saupfercheckweg  1,  69117 Heidelberg, Germany}
 
\author{Jisuke~Kubo}
\email{kubo@mpi-hd.mpg.de}
\affiliation{Max-Planck-Institut f\"ur   Kernphysik  (MPIK), Saupfercheckweg  1,  69117  Heidelberg,  Germany}
\affiliation{Department  of  Physics,  University of Toyama, 3190  Gofuku,  Toyama  930-8555,  Japan}

\author{Manfred~Lindner}
\email{lindner@mpi-hd.mpg.de}
\affiliation{Max-Planck-Institut f\"ur Kernphysik  (MPIK), Saupfercheckweg  1,  69117 Heidelberg, Germany}


\title{\Large Semi-secretly interacting Axion-like particle as an explanation \\\vspace{0.03 in} of Fermilab muon $g-2$ measurement 
}

\begin{abstract}
The muon anomalous magnetic moment measurement has, for more than a decade, been a long-standing anomaly hinting the physics beyond the Standard Model (BSM). The recently announced results from muon $g-2$ collaboration, corresponding to 3.3$\sigma$ deviation from Standard Model value (4.2$\sigma$ in combination with previous measurement) are strengthening the need for new physics coupled to muons. In this letter, we propose a novel scenario in which Standard Model (SM) is augmented by an axion-like particle (ALP) and vector-like fermions. We find that such a model admits an excellent interpretation of recent muon $g-2$ measurement through quantum process featuring ALP interacting with muons and newly introduced fermions. Previously proposed explanations with ALPs utilize interactions with photons and/or SM fermions. Therefore, in this letter we complement and extend such scenarios. We also discuss collider prospects for the model as well as the possibility that ALP is long lived or stable dark matter (DM) candidate.
\end{abstract}

\maketitle

\textbf{\emph{Introduction}.--} 
\noindent
Since the Higgs boson discovery in 2012, high-energy community has been chiefly oriented in the direction of new physics, whose existence was previously unambiguously confirmed only in neutrino oscillation experiments. While LHC has not discovered new particles,  several anomalies appeared at a statistical significance with which conclusive statements can not be established; however such findings still allow us to speculate and hope that new physics is hiding just around the corner. One of the measurements which has previously induced great attention is anomalous muon magnetic moment measurement at BNL that deviates by $\sim 3\sigma$ \cite{Bennett:2006fi,Keshavarzi:2018mgv} from SM. The BNL Muon $g-2$ collaboration has measured it to be $a_\mu({\rm BNL})=116 592 089(63)\times 10^{-11}$  \cite{Bennett:2006fi}, while theoretical predictions find it to be $a_\mu ({\rm theory})=116 591 810(43)\times 10^{-11}$ \cite{Aoyama:2020ynm}. This result was not challenged by any other measurement for years; however, recently results from the $g-2$ experiment at Fermilab were reported \cite{Abi:2021gix}. This announcement has further strengthened the hint for new physics because significant deviation from SM value, corresponding to 3.3$\sigma$ ($4.2\sigma$ in combinaiton with BNL result), was reported. It is indeed important to mention that  community put huge effort in these SM predictions over the two decades and these are listed in chronological order as: HMNT06 \cite{Hagiwara:2006jt}, DHMZ10 \cite{Davier:2010nc}, JS11 \cite{Jegerlehner:2011ti}, HLMNT11 \cite{Hagiwara:2011af}, DHMZ17 \cite{Davier:2017zfy}, KNT18 \cite{Keshavarzi:2018mgv}, see also the recent ones \cite{Aoyama:2020ynm,Davier:2017zfy,Keshavarzi:2018mgv,Colangelo:2018mtw,Hoferichter:2019mqg,Davier:2019can,Keshavarzi:2019abf,Kurz:2014wya,Melnikov:2003xd,Masjuan:2017tvw,Colangelo:2017fiz,Hoferichter:2018kwz,Gerardin:2019vio,Bijnens:2019ghy,Colangelo:2019uex,Colangelo:2014qya,Blum:2019ugy,Aoyama:2012wk,Czarnecki:2002nt,Gnendiger:2013pva,Borsanyi:2020mff}.
   
The letter is based on this result. Our main goal is to scrutinize a previously unconsidered new physics model for which we will show that it can explain the discrepancy between the measurement and the theoretical SM value of muon $g-2$. The cornerstone of our model is the presence of axion-like particle (ALP). Axions are one of the most scrutinized extensions of the SM, being chiefly considered as a solution to the strong CP problem \cite{Peccei:2006as}; note that axions can, however, also serve as a viable DM candidate \cite{Preskill:1982cy,Dine:1982ah,Abbott:1982af,Arias:2012az,Duffy:2009ig}. ALPs are a generalization of axions where their mass and the decay constant are assumed not to be related and are typically treated as independent parameters. 

In the context of an additional contribution to anomalous muon magnetic moment, ALPs have been previously explored in several publications \cite{Marciano:2016yhf, Bauer:2019gfk, Bauer:2017ris, deNiverville:2018hrc} (for other light scalar scenarios, see \cite{Davoudiasl:2018fbb, Jana:2020pxx, Liu:2018xkx, Dutta:2020scq, Cornella:2019uxs, Abdallah:2020vgg, Ghosh:2020tfq, Endo:2020mev, Iguro:2020rby}). In Ref.~\cite{Marciano:2016yhf}, $g-2$ discrepancy is addressed in the model with both ALP-photon and ALP-fermion interactions, through one and two-loop diagrams. In Refs.~\cite{Bauer:2019gfk,Bauer:2017ris}, the authors went one step beyond  and considered the dominant effect of axion-muon-fermion coupling where the fermion is either electron or tau lepton. It is important to mention that one can not achieve correct sign and strength for the muon $g-2$ with just lepton flavor conserving couplings of the ALP. Further, the interpretation with lepton flavor violating couplings suffers from stringent lepton flavor violation constraints \cite{Bauer:2019gfk} and it can be further narrowed down by several low energy experiments. Motivated by this, in this letter, we generalize such approach and consider heavy fermions which are part of the heavy fermion sector. As a result, we get an additional chiral enhancement in the total muon $g-2$ contribution and this interpretation does not lead to tight lepton flavor violating observable. Our model does not feature further extensions of scalar and gauge sectors, in contrast with Ref.~\cite{deNiverville:2018hrc} where the authors considered ALP interactions with dark photon to explain muon $g-2$ discrepancy. 

In what follows, we will first introduce particle physics model under consideration. Then, we will present how Fermilab $g-2$ measurement can be explained in such framework. Before concluding, we will discuss further probes of the considered model, particularly at LHC, and also refer to the possibility 
of ALP being a long lived or stable particle, case in which it can serve as a DM candidate.\\     

\textbf{\emph{The Model and contribution to muon $g-2$}.--} 

\begin{figure}[b!]
   \centering
    \includegraphics[width=0.5\textwidth]{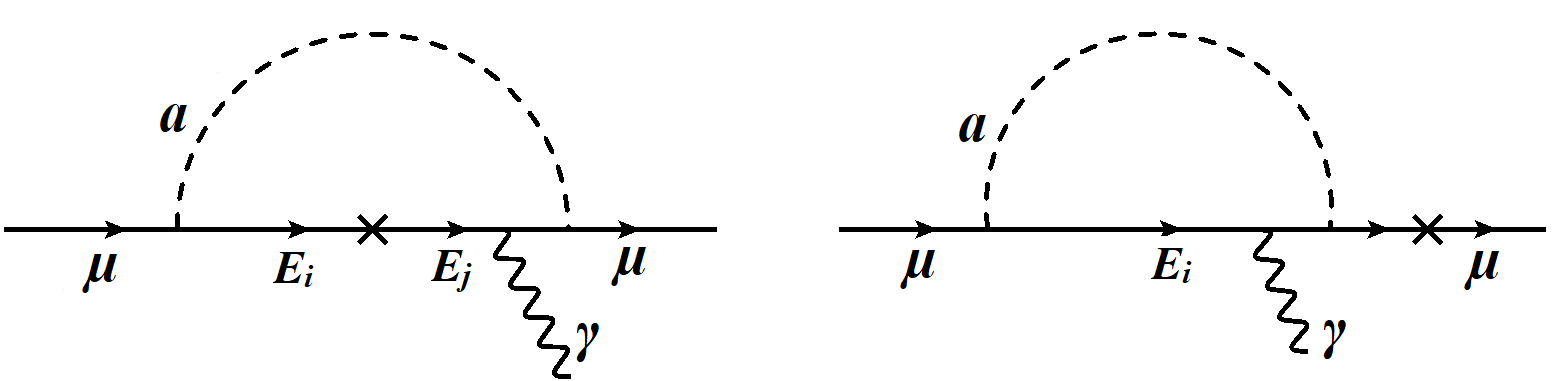}
    \caption{Feynman diagrams representing leading BSM contribution to muon magnetic moment for the considered model. Here $i,j=1,2$ and $i\neq j$. }
    \label{fig:1}
\end{figure}
The relevant part of the Lagrangian containing new fields reads
\begin{align}
\mathcal{L} &= \mathcal{L}_\text{ALP} + 
\mathcal{L}_\text{mass} + \mathcal{L}_\text{yukawa} 
\supset \nonumber \\ &
\left( \frac{C_{\mu E_1}}{2 \Lambda} \, \partial_\alpha a\, \bar{\mu} \gamma^\alpha \gamma_5 E_1 +
\frac{C_{\mu E_2}}{2 \Lambda}\, \partial_\alpha a\, \bar{\mu} \gamma^\alpha \gamma_5 E_2+  \text{h.c.} \right) \,+
 \nonumber \\ &
 \left( m_1 \bar{L}_L L_R + m_2 \bar{E}_{2L} E_{2R} + m_3 \bar{\mu}_R E_{2L} + \text{h.c.} \right)+
  \nonumber \\ &
 (  
y_1 \bar{l}_L H E_{2R} + y_2 \bar{L}_L H \mu_R + y_3 \bar{L}_L H E_{2R} + \nonumber \\ &  y_4 \bar{L}_R H E_{2L} + \text{h.c.} 
 )\,.
\label{eq:lagrangian}
\end{align}
Here, in last three parentheses we separated effective Lagrangian including ALP field, $a$, mass and Yukawa terms. The effective part of the Lagrangian respects $SU(3)_C\times U(1)_Q$.
Higgs field is denoted by $H$, $l_L=(\nu_{\mu L}\,\, \mu_L)^T\sim (1,2,-1)$, $\mu_R \sim (1,1,-2)$ 
are SM fermion SU(2) doublet and singlet, respectively. 
\begin{figure}[htb!]
  \centering
  \includegraphics[width=0.45\textwidth]{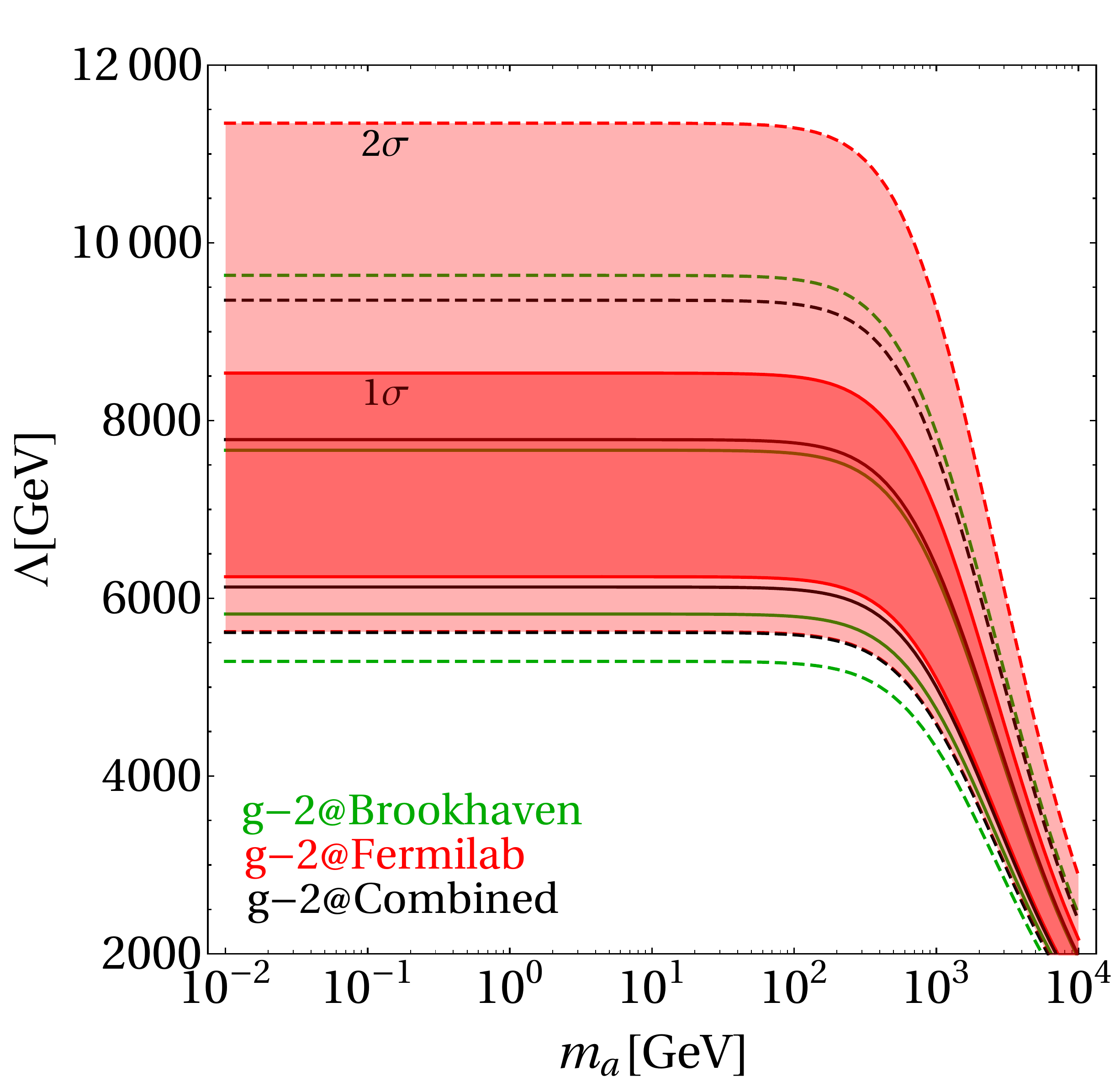}
  \caption{Muon $g-2$ contribution that explains recent measurement at Fermilab (red) in $m_a-\Lambda$ parameter space. For completeness we also indicate region favored by BNL result (green) and combined result (black).}
  \label{fig:2}
\end{figure}

The heavy fermion sector contains $L_L =(N_L \,\, E_{1L})^T\sim (1,2,-1)$, $E_{2L}\sim (1,1,-2)$ together with corresponding fields with right-handed chirality, $L_R$ and $E_{2R}$, respectively. The heavy fermions are hence vector-like and they are assumed to interact only with the second generation of SM leptons. Their masses, $m_1$ and $m_2$ are for simplicity assumed to be degenerate and in what follows will be jointly denoted with $m_E$.
The last term in the second and the first two terms in the third row of Eq. (\ref{eq:lagrangian}) contribute to the muon-heavy fermion mixing. Such interactions are not relevant for muon $g-2$ explanation that we discuss below; however, the induced mixing could set the constraint on the mass scale of heavy fermions as well as induce unwanted ALP coupling to photons. We note that these terms can be forbidden if one invokes parity symmetry under which heavy   fermions and ALP have $(-)$ and SM fields including muons have $(+)$ charge.
The presence of such symmetry also prevents ALP from decaying and makes it a successful DM candidate. We remind the reader that DM production in this scenario would occur through the misalignment mechanism that is independent of any interactions appearing in Eq. (\ref{eq:lagrangian}).

The Feynman diagram illustrating leading BSM contribution to muon $g-2$ is 
shown in Fig. (\ref{fig:1}). We obtained 

\begin{widetext}
\begin{equation}
\Delta a_\mu = \Delta \left(\frac{g-2}{2}\right)= \frac{-q_{E} m_{\mu}^2 m_E^2}{4 \pi^{2}\Lambda^2 } \int_{0}^{1} \mathrm{~d} x \frac{[C_{\mu E_1} C_{\mu E_1}^* +C_{\mu E_2} C_{\mu E_2}^*](x^{2}-x^{3})-Re[C_{\mu E_1} C_{\mu E_2}^* \lambda]\frac{ v_{ew}}{m_{\mu}} x^{2}}{m_{\mu}^{2} x^{2}+\left(m_{E}^{2}-m_{\mu}^{2}\right) x+m_{\mathrm{a}}^{2}(1-x)}\,,
\label{eq:g-2}
\end{equation}
\end{widetext}

where $q_E=1$ is the electric charge of vector-like fermions in units of $e$, $\lambda = \frac{1}{ \sqrt{2}} (y_3 + y_4) $ and $v_{ew} = 246 $ GeV. Let us point our that interactions in the first and the last row of Eq.~(\ref{eq:lagrangian}) contribute to the $g-2$, while the terms in the second row of Eq.~(\ref{eq:lagrangian}) are crucial for making propagating new fermions massive; 
if their mass was vanishing or very small with respect to the electroweak scale, not only would we lose the effect on muon $g-2$,
but, in addition, such model, with light charged states would already be heavily disfavored.

The $g-2$ collaboration has announced \cite{Abi:2021gix} $a_\mu^{exp}-a_\mu^{SM}= (22.9 \pm 6.9)\times 10^{-10}$ ( $(25.1\pm 5.9)\times 10^{-10}$ in combination with BNL measurement). 
In Fig.~(\ref{fig:2}), we indicate the parameter space in our model that explains the reported excess (red) as well the parameter space in which BNL measurement \cite{Bennett:2006fi} (green) and combined result (black) can be addressed. Being unbiased toward the ALP mass, we show results across orders of magnitude in that parameter; on the y-axis we show $\Lambda$. In calculations, we fixed couplings $\lambda=1$ and $C_{\mu E_1}=C_{\mu E_2}=1/\sqrt{2}$, as well as $m_E= 1$ TeV. We note, however, that for small ALP masses, the expression is independent of $m_E$.
In that regime, value of $m_a$ is also not relevant for $\Delta a_\mu$ (see Fig.~(\ref{fig:2})). This is because,
in such case, expression in Eq.~(\ref{eq:g-2}) is simplified to $\Delta a_\mu \propto (m_\mu/\Lambda)^2\, (v_{ew}/m_\mu)$. We wish to stress that the dominant contribution arises from the left diagram in Fig. (\ref{fig:1}) where both $C_{\mu E_1}$ and $C_{\mu E_2}$ enter; this allows us to regulate the sign of $\Delta a_\mu$
and we can explain the recent measurement in case these two couplings (as well as $\lambda$) are of the same sign which for definiteness we take as positive. 
In the parameter space where $m_a>m_E$, the denominator in 
Eq.~(\ref{eq:g-2}) increases which needs to be compensated by smaller value of $\Lambda$, as visible from the figure. We conclude that, 
with natural $\mathcal{O}(1)$ values of all involved couplings and Wilson coefficients, $\Lambda$ should attain values around $\mathcal{O}(10^3-10^4)$ GeV in order to explain the reported excess within the present scenario.

\textbf{\emph{Other probes and constraints}.--} 
\begin{figure}[h]
  \centering
  \includegraphics[width=0.30\textwidth]{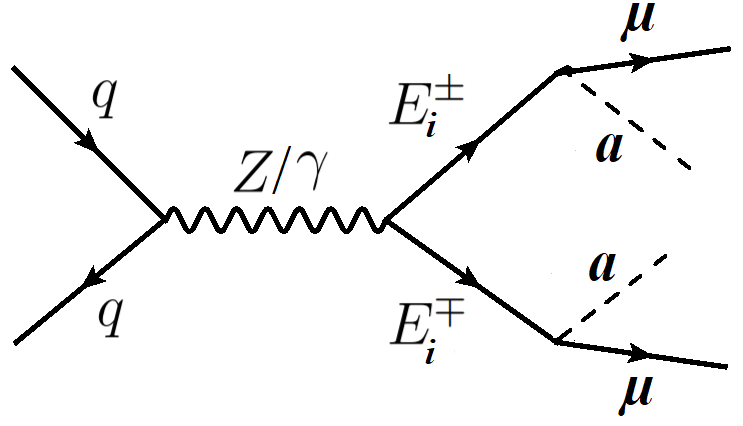}
  \caption{Collider test of the considered model.}
  \label{fig:3}
\end{figure}
Generally, if there is a BSM contribution to muon $g-2$, there are also tight lepton flavor violation constraints that need to be respected, see for instance \cite{Lindner:2016bgg,Miller:2007kk}. In the context of ALPs, the authors of \cite{Bauer:2019gfk} have shown that for that reason, diagonal ALP couplings to fermions 
need to be suppressed; in other words, the off-diagonal ALP coupling to fermions should dominate in order to have a successful, unconstrained contribution to $g-2$ that explains measurements. In this letter, we took a similar path and assumed that vector-like fermions prefer interacting with muons. Such a scenario can be accomplished by utilizing flavor symmetries.
 Further, ALP coupling to photon in this model needs to be suppressed as well since, otherwise, the contribution from diagrams in Ref.~\cite{Marciano:2016yhf} could give competitive or even dominant contribution with respect to the diagram in Fig.~(\ref{fig:1}). 
As previously discussed, mixing between muons and heavy fermions can be avoided with parity symmetry, otherwise presence of such mixing would radiatively induce ALP-photon coupling.

Another constraint on our heavy vector-like fermions comes from the Higgs naturalness; these fermions will contribute to the Higgs mass term $\mu_H^2 H^\dag H$ at the one-loop level \cite{Vissani:1997ys,Clarke:2015gwa,Fabbrichesi:2015zna}, \emph{i.e.}
$|\Delta \mu_H^2| \sim \,(1/4 \pi^2) (2|y_3|^2+2|y_4|^2+ y_3 y_4^*+y_3^* y_4) \,m_E^2 < \,\mu_H^2= m_H^2/2$, where $m_H\simeq 125$ GeV is the Higgs mass.
Since the sign of the radiative correction $\Delta \mu_H^2$ can be controlled by finite terms, we can employ this correction to radiatively generate the desired Higgs mass term, in a similar way as in the case of  the ``neutrino option'' \cite{Brivio:2017dfq,Brdar:2018vjq}. \\
 \begin{figure}
  \centering
  \includegraphics[width=0.48\textwidth]{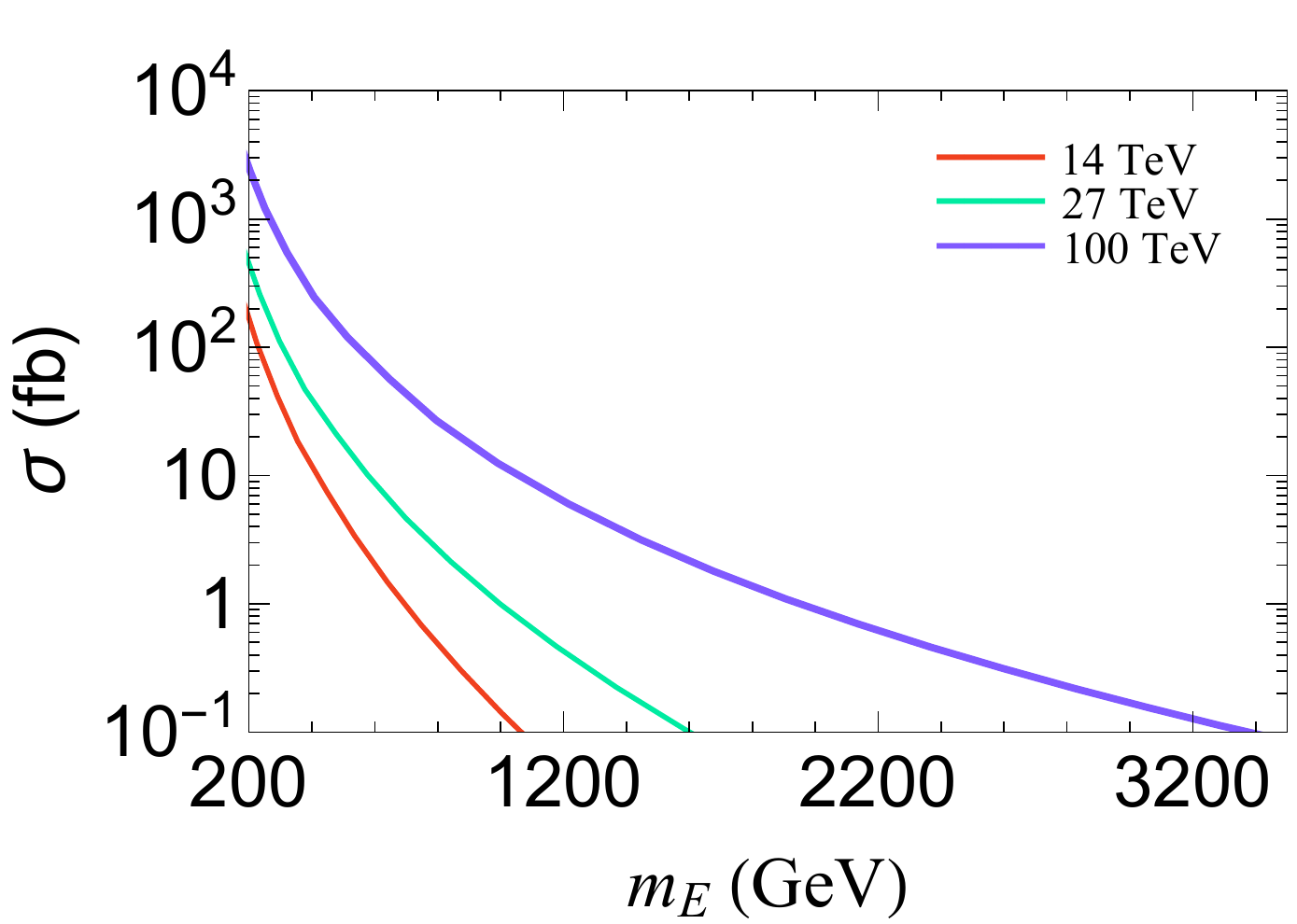}
  \caption{Production cross-sections for the process $pp \to \mu^+ \mu^- + 2 a$ for different centre of mass energies at the LHC.}
  \label{fig:4}
\end{figure}

In Fig.~(\ref{fig:3}), we show a diagram which demonstrates that this scenario can be tested at present and future hadron colliders
(see related studies connecting muon g-2 explanations and collider searches \cite{Calibbi:2018rzv,Ghosh:2021jeg,Chen:2021rnl,Yin:2020afe}). The charged vector-like leptons can be pair-produced at the LHC by standard Drell-Yan processes mediated by $s$-channel $Z/\gamma$ exchange and will further decay dominantly to ALP and muon. To estimate the signal cross-section, we implement our model file in {\tt FeynRules} package~\cite{Christensen:2008py}, and then analyzed the cross section for the signal  using {\tt MadGraph5aMC@NLO}~\cite{Alwall:2014hca} event generator at the parton level. In Fig.~\ref{fig:4}, we show the pair production cross-sections of ALP in association with two muons for different centre of mass energies (14, 27 and 100 TeV) at the LHC.  If ALP is long lived particle, the signature would be dimuon+$\slashed{E_T}$. Otherwise, if ALP can decay within the detector volume, there is a possibility for multi-lepton final state which is a very clean signature; it can essentially be free from SM backgrounds with proper $Z-$ veto and other acceptance cuts. Hence, for our scenario, such search at colliders complements the opportunities at $g-2$ experiments. \\

\begin{figure}
  \centering
  \includegraphics[width=0.45\textwidth]{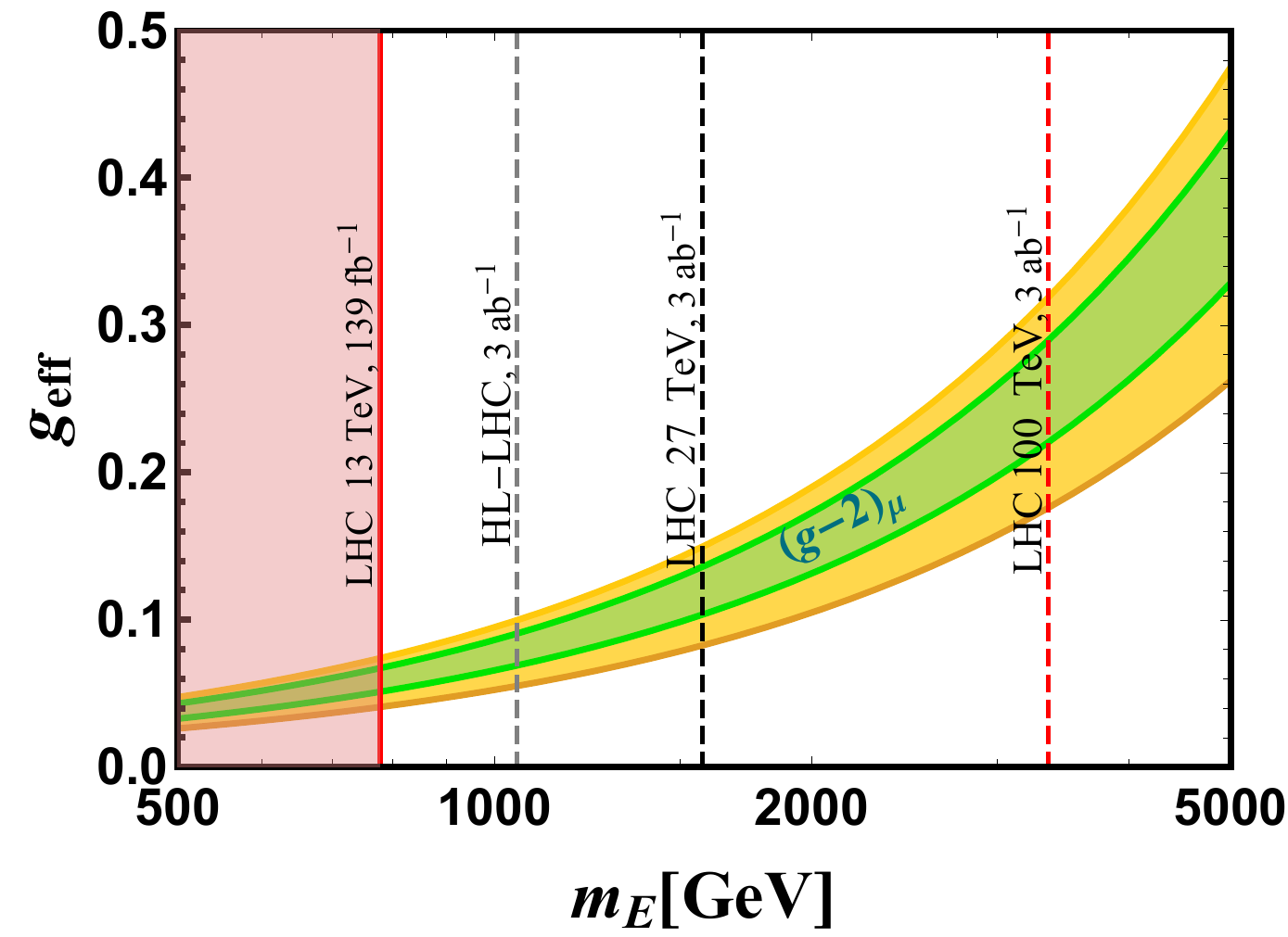}
  \caption{Summary plot in in the $g_{eff}$-$m_{E}$ plane, where effective coupling is defined as $g_{eff}= \mid C_{\mu E_1}C_{\mu E_2}\lambda \mid ({m_E^2}/{\Lambda^2})$. The green and yellow shaded region denote $1\sigma$  and $2\sigma$  allowed range for the muon anomalous magnetic moment measurement \cite{Abi:2021gix} at Fermilab. Red shaded zone is excluded from current $pp \to \mu^+ \mu^- + {E\!\!\!\!/}_{T} $ searches \cite{Aad:2019vnb} at the LHC. Gray, black and red dashed lines indicate the future sensitivity at the HL-LHC, HE-LHC and 100 TeV collider looking at  $pp \to \mu^+ \mu^- + {E\!\!\!\!/}_{T} $ signature. Here we set $m_a=$ 0.1 GeV. See text for details.}
  \label{fig:5}
\end{figure}

For completeness,  in Fig.~\ref{fig:5}, we show the parameter space, in the effective coupling ($g_{eff}$) -- vector-like lepton mass ($m_E$) plane, which can naturally explain  the correct strength and sign of  muon anomalous magnetic moment. Here we define the effective coupling as $g_{eff}= \mid C_{\mu E_1}C_{\mu E_2}\lambda \mid ({m_E^2}/{\Lambda^2})$ and set the ALP mass to 0.1 GeV. The green and yellow shaded regions indicate $1\sigma$  and $2\sigma$  allowed range for the muon anomalous magnetic moment measurement \cite{Abi:2021gix} at Fermilab. Note that the result in Fig.~\ref{fig:5} is still valid for any other ALP masses as long as $m_E > m_a.$ As previously mentioned, after being pair-produced at the LHC by standard Drell-Yan processes, the charged vector-like leptons will further decay dominantly to ALP and muon. However, for our benchmark scenario, we find that the ALP has lifetime long enough to leave the detector, and hence it leads to  novel $pp \to \mu^+ \mu^- + {E\!\!\!\!/}_{T} $  signature at the LHC.  This signature can exactly mimic chargino or slepton (smuon) signature in supersymmetry where chargino/slepton decays back to  charged lepton and long-lived particle with/without neutrinos. Since there are several dedicated searches looking for $pp \to \mu^+ \mu^- + {E\!\!\!\!/}_{T} $ signature \cite{Aad:2019vnb} at the LHC, in this way, stringent  limits on vector-like lepton masses can be set. To derive bound on vector-like lepton mass, we first recast the limit from  current $pp \to \mu^+ \mu^- + {E\!\!\!\!/}_{T} $ searches \cite{Aad:2019vnb} which is based on integrated luminosity of 139 fb$^{-1}$ and center of mass energy $\sqrt{s}=13$ TeV. After comparing the cross sections in \cite{Aad:2019vnb} and in our sceanrio, the limit from  \cite{Aad:2019vnb} can be rescaled. We find that a vector-like lepton of mass up to 780 GeV is excluded in our scenario, see Fig.~\ref{fig:5}. We also extrapolate this result to get the discovery reach at the HL-LHC with a target luminosity of 3 ab$^{-1}$.  We find that a vector-like lepton of mass up to 1049 GeV can be probed in this scenario. We also perform a naive estimation for the sensitivity at future $pp$ collider experiments corresponding to center  of mass energies 27 TeV and 100 TeV by requiring 300 signal events.  In Fig.~\ref{fig:5}, the gray, black and red dashed lines indicate the future sensitivity at the HL-LHC, HE-LHC and 100 TeV collider looking at  $pp \to \mu^+ \mu^- + {E\!\!\!\!/}_{T} $ signature. In summary, LHC is already testing a certain portion of parameter space; however, $m_E\gtrsim 1$ TeV may still be employed to explain Fermilab g-2 result. Future colliders will be able to dramatically improve the sensitivity on the scenario presented in this paper.

\textbf{\emph{Conclusions}.--} 
 The main motivation for this work is recent anomalous muon magnetic moment measurement at Fermilab site, where an excess with respect to the SM was reported. We propose a novel scenario in order to explain this result. The cornerstone of the considered model is ALP. The previous muon $g-2$ explanations with ALPs, considered in the context of BNL measurement, feature ALP interactions with photons and SM fermions. In this paper we complement and extend such framework by scrutinizing the model where ALP interacts semi-secretly, with SM muons and heavy vector-like leptons. We find a viable parameter space in which recent result can be successfully explained. As a complementary probe, we have also discussed collider probes of the model as well as the possibility to have ALP as a viable DM candidate.

\textbf{\emph{Acknowledgements}.--} 
Fermilab is operated by the Fermi Research Alliance, LLC under contract No. DE-AC02-07CH11359 with the United States Department of Energy. J.~K.~is partially supported by the Grant-in-Aid for Scientific Research (C) from the Japan Society for Promotion of Science (Grant No.19K03844).

\bibliography{refs}

\end{document}